\documentclass[11pt,reqno]{amsart}
\usepackage{amssymb,graphicx}

\usepackage{amsfonts,amscd,amsthm,amsmath}
\usepackage{cite}

\begin{document}

\begin{center}
\textbf{\large{Proof of the Borwein-Broadhurst conjecture\\ for a dilogarithmic integral arising in quantum field theory}}

\bigskip
\textbf{Djurdje Cvijovi\'{c}}
\medskip

{\it Atomic Physics Laboratory, Vin\v{c}a Institute of
Nuclear Sciences \\
P.O. Box $522,$ $11001$ Belgrade$,$ Republic of Serbia}\\

\textbf{E-Mail: djurdje@vinca.rs}\\

\vskip 0.5cm

\textbf{Abstract}

\end{center}

\begin{quotation}
Borwein and Broadhurst, using experimental-mathematics techniques, in $1998$ identified numerous hyperbolic 3-manifolds whose volumes are rationally related to values of various Dirichlet L series $\textup{L}_{d}(s)$. In particular, in the simplest case of an ideal tetrahedron in hyperbolic space, they conjectured that a dilogarithmic integral representing the volume equals to $\textup{L}_{-7}(2)$. Here we have provided a formal proof of this conjecture which has been recently  numerically verified (to at least $19,995$ digits, using 45 minutes on 1024 processors) in cutting-edge computing experiments. The proof essentially relies on the results of Zagier on the formula for the value of Dedekind zeta function $\zeta_{\mathbb{K}}(2)$  for an arbitrary field $\mathbb{K}$.
\end{quotation}

\medskip
\noindent {\bf 2008 PACS number(s):} 02.40.-k, 02.10.De, 02.30.Gp, 02.70.-c


\medskip

\section{\bf Introduction}

The following
\begin{equation}\label{eq:1}
I_{7}= \frac{24}{7 \sqrt7}\int_{\pi/3}^{\pi/2}\ln\left(\left|\frac{\tan(\theta) + \sqrt7}{\tan(\theta)-\sqrt7}\right|\right)d\theta
\end{equation}
\noindent and other numerous related integrals arose out of some studies in quantum field theory, in analysis of hyperbolic manifolds whose complementary volumes result from evaluations of Feynman diagrams \cite{Borwein, Broadhurst,Broadhurst1, Broadhurst2,Lunev}. $I_{7}$ represents the volume of an ideal tetrahedron in hyperbolic space $\mathbb{H}_{3}$ and is the simplest of $998$ {\em empirically} ({\em i.e.} by using various experimental-mathematics techniques) determined cases where the volume of a closed hyperbolic 3-manifold is rational multiple of values of various Dirichlet L series \cite{Broadhurst}. Here an ideal (or totally asymptotic) tetrahedron in $\mathbb{H}_{3}$ is a hyperbolic tetrahedron  with all four vertices at infinity.

In 1998 Borwein and Broadhurst conjectured that
\begin{equation}\label{eq:2}
I_{7} \mathop  = \limits^? \textup{L}_{-7}(2)\cong 1.151925470544491
\end{equation}
\noindent where $\textup{L}_{-7}(s)$ is the primitive Dirichlet L-series modulo seven \cite{Broadhurst, Borwein1, Bailey0, Bailey1, Bailey,Bailey2}. The sign ? here indicates that numerical verification of this "identity" has been performed, but that no formal proof of it is yet known \cite{Bailey2}. The verification of (\ref{eq:2}) has been performed on several occasions, and recently, the agreement  between the values for $I_{7}$  and $\textup{L}_{-7}(s)$ to at least $19,995$ digits, using 45 minutes on 1024 processors has been found in computations at the very edge of presently available numerical techniques and computing technology (see \cite{Bailey3} and Remark 1 below). Note that

\begin{equation}\label{eq:3}
\phi_{7}= \textup{tan}^{-1}\left(\sqrt7\right)\cong 1.209429202888189
\end{equation}
is the only (and particularly "nasty") singularity of the integrand in the integral (\ref{eq:1}) inside the interval $(\tfrac{\pi}{3},\tfrac{\pi}{2})$.

\begin{figure}[htp]
\centering
\includegraphics{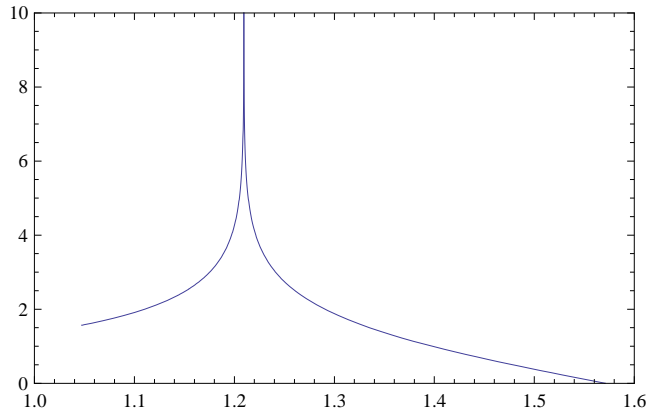}
\caption{Plot of integrand function with singularity}
\end{figure}

In this paper our aim is to rigorously demonstrate  the truth of the Borwein-Broadhurst conjecture (\ref{eq:2}). In order to do that, first we evaluate the integral $I_{7}$ in a closed form as follows

\begin{align}\label{eq:4}
I_{7} =  \frac{4}{7 \sqrt7} &\left[3 \textup{Cl}_2(2\phi_{7})-3 \textup{Cl}_2(4\phi_{7}) +  \textup{Cl}_2(6\phi_{7})\right]
\end{align}
\noindent in terms of the values of the Clausen function  $\textup{Cl}_2(\theta)$.

Second, we show that

\begin{align}\label{eq:5}
\textup{L}_{-7}(2)= \frac{2}{\sqrt7} \left[\textup{Cl}_2\left(\frac{2 \pi}{7}\right)+ \textup{Cl}_2\left(\frac{4\pi}{7}\right)- \textup{Cl}_2\left(\frac{6\pi}{7}\right)\right].
\end{align}

\noindent Third, a formal proof is given for the following striking and unexpected relation (verified numerically to $1,800$-digit accuracy \cite{Broadhurst}) between six values of $\textup{Cl}_2(\theta)$

\begin{align}\label{eq:6}
& 2 \left[3 \textup{Cl}_2(2\phi_{7})- 3 \textup{Cl}_2(4\phi_{7}) + \textup{Cl}_2(6\phi_{7})\right]\nonumber
\\
\mathop  = \limits^? &7  \left[\textup{Cl}_2\left(\frac{2\pi}{7}\right) +\textup{Cl}_2\left(\frac{4\pi}{7}\right)
-\textup{Cl}_2\left(\frac{6\pi}{7}\right)\right].
\end{align}

\medskip
The conjecture (\ref{eq:2}) is a rather interesting illustration of current work and results in  {\it experimental mathematics} \cite{Bailey0, Bailey1, Bailey, Borwein1, Bailey2}, namely a type of mathematical investigation that stresses the importance and significance of computational (or "numerical") experiments, and, in which, advanced computing technology is used to explore mathematical structures, test conjectures and suggest generalizations. Computations can in many cases provide very compelling evidence for mathematical assertions, however, results discovered experimentally will, in general, lack some of the rigor associated with mathematics but will provide general insights into mathematical problems to guide any further exploration, either experimental or traditional. As in experimental science, experimental mathematics can be used to make predictions which can  be verified or falsified on the bases of additional experiments. Some examples of research tools are  computer and symbolic algebra,  arbitrary precision arithmetic, Gr\"obner bases, integer relation algorithms (such as the LLL algorithm and PSLQ algorithm), computer visualization, cellular automata and related structures, and various databases. A significant milestone and achievement of experimental mathematics was the discovery in 1995 of the Bailey–-Borwein–-Plouffe formula for the binary digits of $\pi$. For the sake of an illustration, we mention several recent papers \cite{Bailey4, Boukraa, Boukraa1} as an example of power of this technique in the context of physics-related problems.

\medskip
\section{\bf Preliminaries}

Before proceeding to a proof of (\ref{eq:2}) we provide some preliminaries. Let $\left(\frac{d}{n}\right)$ be the Kronecker symbol defined for a positive integer $n$ and a non-square integer $d$ satisfying the congruences $d\equiv 0$ or $1$ (mod 4) (some of the admissible values are $-8, -7, -4, -3, 5, 8, 12, 13, 17, 20, 21$), then, for  $\mathfrak{R}(s) >1 $, define Dirichlet L series $\textup{L}_d(s)$ in the following way (for more details, see, for instance, \cite{Ayoub, Landau})
\begin{equation}\label{eq:7}
\text{L}_{d}(s)= \sum_{n=\,1}^{\infty}\left(\frac{d}{n}\right)\,\frac{1}{n^s}.
\end{equation}

\noindent Symbol $\left(\frac{d}{n}\right)$, for a given admissible $d$, only assumes values $1, -1$ and $0$, and is a periodic function with a period of $|d|$, thus $\textup{L}_{d}(s)$ can be rewritten as the following sum
\begin{equation*}
\textup{L}_{d}(s)= \frac{1}{|d|^s} \sum_{\ell =\,1}^{|d|-1}\left(\frac{d}{\ell}\right)\,\zeta\left(s,\frac{\ell}{|d|}\right),
\end{equation*}
\noindent where $\zeta(s,a)$ denotes the Hurwitz  zeta function
\begin{equation*}
\zeta \left(s, a\right)=\sum_{m = 0}^{\infty}\frac{1}{(m + a)^s},
\end{equation*}
\noindent $(\mathfrak{R}(s)>1;\, a \not\in \left\{0, -1, -2, \ldots\right\})$. Note that $\zeta(s): = \zeta(s,1)$ is the Riemann zeta function.

Now, since $\left(\frac{-7}{n}\right),$ $n\in\mathbb{N},$ is periodic in $n$ with a period of $7$ and $1,1,-1,1,-1,-1$ and $0$ are its first seven values, we have

\begin{align*}\textup{L}_{-7}(s) = \frac{1}{7^s} \left[ \zeta\left(s,\frac{1}{7}\right)+ \zeta\left(s,\frac{2}{7}\right)- \zeta\left(s,\frac{3}{7}\right) + \zeta\left(s,\frac{4}{7}\right) - \zeta\left(s,\frac{5}{7}\right) - \zeta\left(s,\frac{6}{7}\right)\right].
\end{align*}

Clearly, $\textup{L}_{-7}(2)$ can be expressed in terms of the Hurwitz zeta function values, however, we have failed to utilize this connection in our attempts to establish (\ref{eq:2}).

\medskip
\noindent{\bf Remark 1.} $\textup{L}_{-7}(2)$ expressed by means of six the $\zeta(2,s)$ values was used by Bailey and Borwein in their tests  of the conjecture (\ref{eq:2}). $I_{7}$ was computed (using highly parallel tanh-sinh quadrature) to $20,000$ digit accuracy and the result was compared with a $20,000$-digit evaluation of the six-term infinite series for $\textup{L}_{-7}(2)$. The evaluation of the integral $I_{7}$ (by System X at Virginia Tech, an Apple G5-based parallel supercomputer comprised of 1,100 2GHz dual processor Power Mac G5 computers) is probably the highest-precision non-trivial numerical integration performed to date.

The Clausen function (of order 2) $\textup{Cl}_2(\theta)$, sometimes also called the Clausen integral, is a real function for all $\theta\in\mathbb{R}$ and is given by (\cite[p. 102)]{Lewin2}
\begin{align}\label{eq:8}
\textup{Cl}_2(\theta)= \sum_{m=\,1}^{\infty} \frac{\sin\left(m\theta\right)}{m^2}=-\int_{0}^{\theta}\ln\left|2 \sin \left(\frac{t}{2}\right)\right|dt.
\end{align}
\noindent The standard texts on the theory of $\textup{Cl}_2(\theta)$ (and various related functions) are Lewin's books \cite{Lewin, Lewin2}.

From the definition (\ref{eq:8}) it is not difficult to deduce some  elementary properties of $\textup{Cl}_2(\theta)$.

\vskip 5 mm
\noindent{\bf Lemma 1.} We have:
\begin{align*}
&\textup{(a)}\quad\textup{Cl}_2(-\theta) = - \textup{Cl}_2(\theta);
\\
&\textup{(b)}\quad\textup{Cl}_2(\theta+2 m \pi)= \textup{Cl}_2(\theta),\,\, \text{for}\; m\in\mathbb{Z};\hskip65mm
\\
&\textup{(c)}\quad\textup{Cl}_2(\pi+\theta)= - \textup{Cl}_2(\pi-\theta);
\\
&\textup{(d)}\quad\textup{Cl}_2(m \pi) = 0 ,\,\, \textup{for}\; m\in\mathbb{Z}.
\end{align*}

The following lemma is familiar multiplication formula for the Clausen function  \cite[p. 105]{Lewin2}.

\vskip 5 mm
\noindent{\bf Lemma 2} (Multiplication formula){\bf .} If $m \in \mathbb{N}$, then
\begin{equation*}
\textup{Cl}_2(m\, \theta) = m\, \sum_{\ell=\,0}^{m -1} \textup{Cl}_2\left(\theta + \ell \,\frac{2 \pi}{m}\right).
\end{equation*}

In particular we have the duplication
\begin{align*}
\frac{1}{2}\,\textup{Cl}_2(2\, \theta) & = \textup{Cl}_2(\theta) + \textup{Cl}_2(\pi+\theta)= \textup{Cl}_2(\theta) - \textup{Cl}_2(\pi-\theta)
\end{align*}
\noindent and triplication formula
\begin{align*}
\frac{1}{3}\,\textup{Cl}_2(3\, \theta) = \textup{Cl}_2(\theta) + \textup{Cl}_2\left(\theta + \frac{2 \pi}{3}\right) + \textup{Cl}_2\left(\theta - \frac{2 \pi}{3}\right).
\end{align*}

Both integrals given by Lemma 3 below, in view that by Lemma 1(d) we have $\textup{Cl}_2(0)=0$, are readily available from the integral
\begin{align*}
\int\ln\left[\tan(\theta)+ \tan(\phi)\right]d\theta =  -\ln\left[\cos(\phi)\right]\,\theta- \frac{1}{2}\, \textup{Cl}_2(2 \theta+ 2\phi) - \frac{1}{2}\, \textup{Cl}_2(\pi - 2\theta) + C,
\end{align*}
\noindent which, in turn, can be found in  books by Lewin (see, for instance, \cite[p. 260, Eq. (8.69)]{Lewin2}).

\vskip 5 mm
\noindent{\bf Lemma 3.} In terms of the Clausen function defined as in (\ref{eq:8}), we have:

\begin{align*}
\textup{(a)}\quad&\int_{\phi}^{x}\ln\left(\frac{\tan(\theta) + \tan(\phi)}{\tan(\theta)-\tan(\phi)}\right)d\theta = \frac{1}{2}\,\textup{Cl}_2(4 \phi) - \frac{1}{2} \, \textup{Cl}_2(2 x + 2 \phi)+ \frac{1}{2} \,\textup{Cl}_2(2 x - 2 \phi);
\\
\textup{(b)}\quad &\int_{x}^{\phi}\ln\left(\frac{\tan(\phi)+\tan(\theta)}{\tan(\phi)-\tan(\theta)}\right)d\theta=- \frac{1}{2}\,\textup{Cl}_2(4 \phi)+\frac{1}{2} \,\textup{Cl}_2(2 x+2 \phi)- \frac{1}{2} \,\textup{Cl}_2(2 x - 2 \phi).
\end{align*}

\bigskip
\section{\bf Formal proof of (2)}

We are now ready to demonstrate that Equations (\ref{eq:4}), (\ref{eq:5}) and (\ref{eq:6}) hold true and in that way formally prove the conjectured identity (\ref{eq:2}).

In order to obtain (\ref{eq:4}) we shall use Lemmas 2 and 3.  Since the integral $I_{7}$ in (\ref{eq:1}) could be decomposed as
\begin{align*}
\frac{7 \sqrt 7}{24}\,I_{7}  =  \int_{\pi/3}^{\phi_{7}}\ln\left(\frac{\tan(\phi_{7}) + \tan(\theta)}{\tan(\phi_{7})-\tan(\theta)}\right)d\theta + \int_{\phi_{7}}^{\pi/2}\ln\left(\frac{\tan(\theta) + \tan(\phi_{7})}{\tan(\theta)-\tan(\phi_{7})}\right)d\theta,
\end{align*}

\noindent  $\phi_{7}$ being the singularity (\ref{eq:3}), by making use of Lemma 3, we find that

\begin{align}\label{eq:9}
\frac{7 \sqrt 7}{24}\,I_{7} =  - \textup{Cl}_2\left(\pi + 2 \phi_{7}\right) + \frac{1}{2} \left[\textup{Cl}_2\left(2 \phi_{7} +\frac{2 \pi}{3}\right)+ \textup{Cl}_2\left(2 \phi_{7} - \frac{2 \pi}{3}\right) \right].
\end{align}

\noindent On the other hand, by Lemma 2, we have
\[
\textup{Cl}_2\left(\pi + 2 \phi_{7}\right)= \frac{1}{2}\, \textup{Cl}_2\left(4 \phi_{7}\right)-\textup{Cl}_2\left(2 \phi_{7}\right)
\]
\noindent and
\begin{align*} \textup{Cl}_2\left(2 \phi_{7} + \frac{2 \pi}{3}\right)+ \textup{Cl}_2\left(2 \phi_{7} - \frac{2 \pi}{3}\right)= \frac{1}{3}\, \textup{Cl}_2\left(6 \phi_{7}\right) -\textup{Cl}_2\left(2 \phi_7\right).
\end{align*}

\noindent In the end, these expressions and (\ref{eq:9}) together give the sought result given by (\ref{eq:4}). Observe that recently Coffey \cite{Coffey} evaluated differently this integral in terms of $\textup{Cl}_2\left(\theta\right)$.

To obtain (\ref{eq:5}), we apply the Fourier series  \cite[p. 221]{Landau}
\begin{equation*}\left(\frac{d}{n}\right)= \frac{1}{\sqrt d}\sum_{\ell=\,1}^{|d|} \left(\frac{d}{\ell}\right)e^{\imath \frac{2 \ell \pi  n}{ |d|}}, \quad(\imath^2:=-1)
\end{equation*}
\noindent which, in the case when $d < 0$, becomes
\begin{equation*}
\left(\frac{d}{n}\right)=\frac{1}{\sqrt{|d|}}\sum_{\ell=1}^{|d|-1}\left(\frac{d}{\ell}\right)\sin\left(\frac{2 \,\ell n \pi}{|d|} \right),
\end{equation*}
\noindent so that the sequence $\left(\frac{-7}{n}\right)$, $n\in\mathbb{N}$, can be expanded as
\begin{equation*}\left(\frac{-7}{n}\right)= \frac{2}{\sqrt7} \left[\sin\left(\frac{2 n \pi}{7}\right)+\sin\left(\frac{4 n \pi}{7}\right)-\sin\left(\frac{6 n \pi}{7}\right)\right].
\end{equation*}
It is obvious then, that, by making use of this expansion and the above definitions of $\textup{L}_{d}(s)$  and $\textup{Cl}_2\left(\theta\right)$, (\ref{eq:7}) and (\ref{eq:8}), we would get the required formula (\ref{eq:5}).

What remains is to provide the most important part of our proof of (\ref{eq:2}), {\em i.e.} to show that (\ref{eq:6}) holds true. It is interesting  that it has not been noticed before that the validity of the relation (\ref{eq:6}) is an immediate consequence of the results of Zagier \cite[p. 287]{Zagier}.

More precisely, (\ref{eq:6}) was essentially proved by Zagier, who deduced the value of $\text{L}_{-7}(2)$  in two different and independent ways, and, as a result, obtained that $\text{L}_{-7}(2)$ equals to
\begin{equation}\label{eq:10a}
\frac{2}{\sqrt{7}}\left[ A\left(\cot\frac{\pi}{7}\right)+ A\left(\cot\frac{2 \pi}{7}\right)+ A\left(\cot\frac{4 \pi}{7}\right)\right]\tag{10a}
\end{equation}
\noindent  and also equals to
\begin{equation}\label{eq:10b}
\frac{12}{7 \sqrt{7}}\left[ 2 A\left(\sqrt 7\right)+ A\left(\sqrt 7+2 \sqrt 3\right)+ A\left(\sqrt 7- 2 \sqrt 3\right)\right],\tag{10b}
\end{equation}
\noindent where $A(x)$ is the real-valued function defined by
\begin{equation*}
A(x)=\int_{0}^{x} \frac{1}{1+t^2} \ln\left(\frac{4}{1+t^2}\right)\,dt\quad(x\in\mathbb{R}).
\end{equation*}

As a matter of fact, in \cite{Zagier}, Zagier  investigates the values of the Dedekind zeta function (of an algebraic field $\mathbb{K}$) $\zeta_{\mathbb{K}}(s)$ at a positive even integral argument $s=2 m$.  He gives a {\it geometric} proof of the formula for the value of  $\zeta_{\mathbb{K}}(2)$ for an arbitrary field $\mathbb{K}$ (Theorem 1) and the proof involves the interpretation of $\zeta_{\mathbb{K}}(2)$  as the volume of a hyperbolic manifold. In addition,  by routine {\it number-theoretical} tools, he also finds the formula for the value of $\zeta_{\mathbb{K}}(2 m)$ valid if $\mathbb{K}$ is abelian over $\mathbb{Q}$ (Theorem 2). The above expressions in (\ref{eq:10a}) and (\ref{eq:10b}) are, respectively, examples of the application of Theorem 2 and Theorem 1 in the case of the  imaginary quadratic  field $\mathbb{Q}(\sqrt{-7})$. Observe that then the Dedekind zeta function factors as $$\zeta_{\mathbb{Q}(\sqrt{-7})}(s)= \zeta (s) \, \text{L}_{-7}(s),$$ where the Riemann zeta function and Dirichlet L series, $\zeta (s)$ and $\textup{L}_{-7}(s)$, are the afore-defined functions. Upon recalling that $\zeta(2)=\frac{\pi^2}{6}$ we have

\begin{equation*}\textup{L}_{-7}(2)=\dfrac{\zeta_{\mathbb{Q}(\sqrt{-7})}(2)}{\zeta(2)}
=\dfrac{6 \, \zeta_{\mathbb{Q}(\sqrt{-7})}(2)}{\pi^2},\end{equation*}

\noindent and it is clear now that the values of $\text{L}_{-7}(2)$ given by (\ref{eq:10a}) and (\ref{eq:10b}) follow from the results for $\zeta_{\mathbb{Q}(\sqrt{-7})}(2)$ found by Zagier and explicitly stated as Equations (5) and (6) in \cite[p. 287]{Zagier}.

To obtain (\ref{eq:6}), we only need to  rewrite the expressions (\ref{eq:10a}) and (\ref{eq:10b}) in terms of the Clausen function $\textup{Cl}_2 \left(\theta\right)$ and show that they respectively appear on the left-hand and right-hand side of (\ref{eq:6}).

Indeed, from
\begin{align*}\dfrac{d}{dx}\,\textup{Cl}_2 \left(2\, \textup{cot}^{-1}(x)\right)=-\frac{2}{1+x^2}\,\textup{Cl}_2^{'}\left(2\, \textup{cot}^{-1}(x)\right)=\frac{1}{1+x^2}\,\log\left(\frac{4}{1+x^2}\right)
\end{align*}
\noindent we conclude that
\begin{equation}\label{eq:11}
A(x)=\textup{Cl}_2 \left(2\, \textup{cot}^{-1}(x)\right).\tag{11}
\end{equation}

\noindent Next, starting from (\ref{eq:10a}) and by use of (\ref{eq:11}), it is easy to derive our expression for $\textup{L}_{-7}(2)$ in (\ref{eq:5}). It suffices to observe that $\textup{Cl}_2\left(\frac{8 \pi}{7}\right)=-\textup{Cl}_2\left(\frac{6 \pi}{7}\right)$ (see Lemma 1(c)). Similarly, starting from (\ref{eq:10b}), by Lemma 4 below and by the duplication and triplication formulas given by Lemma 2, upon noting that $\phi_{7}= \textup{tan}^{-1}\left(\sqrt7\right)$, we arrive at
\begin{equation}\label{eq:12}
\textup{L}_{-7}(2)= \frac{4}{7 \sqrt{7}}\left[3 \textup{Cl}_2(2\phi_{7})- 3 \textup{Cl}_2(4\phi_{7}) + \textup{Cl}_2(6\phi_{7})\right].\tag{12}
\end{equation}

Finally, it is clear that combining (\ref{eq:5}) and (\ref{eq:12}) leads to the relation between six values of $\textup{Cl}_2(\theta)$ given in (\ref{eq:6}).

In the  derivation of (\ref{eq:12}) we require Lemma 4. Here, the well-known formula \cite[p. 769]{Prudnikov}
\begin{equation*}
\textup{cot}^{-1}(x)+\textup{tan}^{-1}(x)=\frac{\pi}{2}\quad(x\geq 0),
\end{equation*}
\noindent gives Part (a), while Parts (b) and (c) follow from Part (a) and
\begin{equation*}
\textup{cot}^{-1}(\sqrt 7 \pm 2 \sqrt3) + \textup{cot}^{-1}(\sqrt 7) = \pm\frac{\pi}{6}
\end{equation*}
\noindent which can be easily verified by taking the cotangent of both sides and recalling the formula for the cotangent of a sum.

\smallskip
\noindent{\bf Lemma 4.} We have
\begin{align*}\textup{(a)}\quad&\textup{cot}^{-1}(\sqrt 7)=\frac{\pi}{2}-\textup{tan}^{-1}(\sqrt 7);\hskip65mm
\\
\textup{(b)}\quad&\textup{cot}^{-1}(\sqrt 7+2 \sqrt 3)=\textup{tan}^{-1}(\sqrt 7)-\frac{\pi}{3};
\\
\textup{(c)}\quad&\textup{cot}^{-1}(\sqrt 7 - 2 \sqrt 3)=\textup{tan}^{-1}(\sqrt 7)- \frac{2 \pi}{3}.
\end{align*}

To conclude, in this paper, by proving (\ref{eq:4}), (\ref{eq:5}) and (\ref{eq:6}), we have provided a formal proof of the Borwein-Broadhurst conjecture for a dilogarithmic integral arising in quantum field theory.

\bigskip
\section*{\bf Acknowledgements}
The present investigation was supported by the {\it Ministry of Science of the Republic of Serbia} under Research Project  Number 142025.

\end{document}